\documentclass[usegraphicx,useAMS,usenatbib]{mn2e}

\usepackage{amssymb}
\usepackage{times}
\usepackage{aas_macros} 

\usepackage{graphicx}
\def\note #1]{{\bf #1]}}
\def\acir{$\alpha$~Cir}

\def\alphacircini{$\alpha$~Circini}
\def\betacir{$\beta$~Cir}
\def\sigmalup{$\sigma$~Lup}

\def\mcmc{MCMC} 
\def\wire{{\em WIRE}}
\def\vwa{VWA}
\def\cesam{CESAM}
\def\synth{SYNTH}
\def\synthmag{SYNTHMAG}
\def\vald{VALD}
\def\llmodels{LLModels}
\def\atlasni{ATLAS9}
\def\gjstar{GJ~560~B}

\def\teff{$T_{\rm eff}$}
\def\logg{$\log g$}
\def\feh{[Fe/H]}
\def\iet{\,{\sc i}}
\def\ito{\,{\sc ii}}
\def\ione{\,{\sc i}}
\def\itwo{\,{\sc ii}}
\def\itre{\,{\sc iii}}
\def\loggf{$\log gf$}
\def\str{Str\"omgren}
\def\hbeta{$H_\beta$}
\newcommand{\eeee}{El.}
\def\ie{i.e.}
\def\eg{e.g.}
\def\cf{cf.}
\def\rms{{\em rms}}
\def\kms{km\,s$^{-1}$}
\def\hipp{{\em Hipparcos}}
\def\dss{$\delta$~Scuti}

\title[The fundamental parameters of $\alpha$~Cir]
      {The fundamental parameters of the roAp star $\alpha$~Circini\thanks{Based on observations with the Sydney University Stellar Interferometer at the Paul Wild Observatory, Narrabri (Australia) and observations collected at the European Southern Observatory, Paranal (Chile), as part of programmes 072.D-0138 and 077.D-0150.}}

\author[H. Bruntt et. al.]
       {H. Bruntt$^{1}$\thanks{E-mail: bruntt@physics.usyd.edu.au}, 
        J.~R. North$^{1}$,  
        M.~Cunha$^{2}$, 
        I.~M.~Brand{\~a}o$^{2,3}$, 
        V.\ G.\ Elkin$^{4}$, 
        D.~W.~Kurtz$^{4}$,\newauthor
        J. Davis$^{1}$, 
	T.~R. Bedding$^{1}$, 
        A.~P. Jacob$^{1}$, 
        S.~M. Owens$^{1}$,
	J.~G. Robertson$^{1}$, W.~J. Tango$^{1}$, \newauthor
        J.~F. Gameiro$^{2,3}$, M.~J. Ireland$^{5}$ and P.~G. Tuthill$^{1}$\\
	$^1$School of Physics, University of Sydney, NSW 2006, Australia\\
	$^2$Centro de Astrof\'isica da Universidade do Porto, Rua das Estrelas, 4150-Porto, Portugal\\
        $^3$ Departamento de Matem\'atica Aplicada, Faculdade de Ci\^encias, Universidade do Porto, 4169 Porto, Portugal\\
	$^4$Centre for Astrophysics, University of Central Lancashire, Preston PR1~2HE, UK\\
	$^5$Planetary Science, MS 150--21, Caltech, 1200 E.~California~Blvd, Pasadena, CA 91125, USA}

\begin{document}

\date{Accepted 2008 February 25.  Received 2008 February 5; in original form 2007 December 10}

\pagerange{\pageref{firstpage}--\pageref{lastpage}} \pubyear{2008}

\maketitle

\label{firstpage}

\begin{abstract}
We have used the Sydney University Stellar Interferometer (SUSI) to measure the angular diameter of \acir.
This is the first detailed interferometric study of a rapidly oscillating A (roAp) star, 
\acir\ being the brightest member of its class.
We used the new and more accurate \hipp\ parallax to determine the radius to be $1.967\pm0.066\,{\rm R}_\odot$.
We have constrained the bolometric flux from calibrated spectra to determine 
an effective temperature of $7420\pm170$\,K. 
This is the first direct determination of the temperature of an roAp star.
Our temperature is at the low end of previous estimates, 
which span over 1000~K and were 
based on either photometric indices or spectroscopic methods.
In addition, we have analysed two high-quality spectra of \acir, obtained at
different rotational phases and we find evidence for the presence of spots.
In both spectra we find nearly solar abundances of C, O, Si, Ca and Fe, high abundance
of Cr and Mn, while Co, Y, Nd and Eu are overabundant by about 1 dex.
The results reported here provide important observational constraints for 
future studies of the atmospheric structure and pulsation of \acir.
\end{abstract}



\section{Introduction}

The rapidly oscillating Ap (roAp) stars are
chemically peculiar main-sequence stars with effective temperatures 
around $6500$--$8000$\,K. They are found around the classical instability strip 
but have quite different properties from the \dss\ stars, due to their slow rotation
and strong global magnetic fields (typically of several kilogauss (kG), and up to $24$\,kG).
They have peculiar abundances of Sr, Co, 
and certain rare earth elements such as Eu, Nd, Pr, Tb, and Th that 
are found in high concentration high in the atmosphere \citep{ryab07}. 
The roAp stars present an intriguing possibility to study 
element stratification, stellar evolution and pulsation 
in the presence of magnetic fields. 

The roAp stars oscillate in high overtone, low-degree p\,modes 
similar in period to the 5-minute acoustic oscillations in the Sun, 
but with coupling of the magnetic field, 
rotation and the oscillations that must be considered \citep{cunha05}. 
The excitation of high-overtone pulsations, instead of the low overtones found in $\delta$\,Scuti stars, 
is thought to be directly related to the strong magnetic fields, which may suppress envelope convection, 
increasing the efficiency of the opacity mechanism in the region of hydrogen ionization \citep{balmforthetal01,cunha02}.
Asymptotic theory, valid for high overtones, 
predicts that regularly spaced peaks will dominate the frequency spectra of roAp stars, 
and this has been observed in photometric studies of several stars of the class \citep{matt99}. 
More recently, detailed pulsation studies of roAp stars have used high-resolution spectra collected at
high cadence. From ana\-lysis of the rare-earth element lines a complex picture has emerged,
in which the mode amplitudes and sometimes phases depend on atmospheric height,
indicating the presence of running magneto-acoustic waves \citep{kurtz06newtype, ryab07}.

\alphacircini\ (HR~5463, HD~128898; $V=3.2$)
is the brightest of 40 currently known roAp stars. 
It has one dominant oscillation mode with a 
period of $6.8$\,min and a semi-amplitude of 2.5\,mmag in $B$.
Four other modes with amplitudes lower by an order of magnitude 
were detected in an extensive three-site ground-based campaign by \cite{kurtz94}, 
but a clear signature of the large separation has so far not been found in the star.
Recently, the \wire\ satellite observed \acir\ which for the first time
directly showed the rotational modulation ($P_{\rm rot}=4.46$\,d) and
a clear regular structure of the pulsation peaks, with three modes having comparable amplitudes.
The equidistant separation between them has been interpreted 
as half the large separation \citep{bruntt08a}.
This new result offers the possibility to confront observations
with theoretical evolution and pulsation models for the star.

To make progress in the asteroseismic modelling of \acir\ we not only need to
measure accurate frequencies with secure mode identification, but also
to know with good accuracy the fundamental parameters of the star. 
It is the goal of the current paper to determine
the effective temperature, luminosity, mass and chemical composition of \acir.
A core-wing anomaly is seen in the Balmer lines of Ap stars \citep{cowley01},
indicating that the temperature structure is different from normal A-type stars.
Therefore, estimates of fundamental parameters from photometric indices or
spectral analysis are likely to be affected by systematic effects \citep{koch05}. 
For example, \teff\ estimates for \acir\ cover the range $7470 - 8730$\,K 
(considering 1$\sigma$ uncertainties; \cf\ Table~\ref{tab:fund}), 
spanning a large part of the range of \teff\ of the stars belonging to the roAp class.
Also, estimating the surface gravity from a classical abundance analysis (\eg\ \citealt{kupka96})
by requiring lines of neutral and ionized lines 
to yield the same abundance is questionable \citep{ryab02}.

The importance of obtaining accurate fundamental parameters
through interferometry for detailed asteroseismic studies was 
discussed in detail by \cite{creevey07} and \cite{cunha07}.
In the current paper we present new interferometric data to
measure the angular diameter of \acir. We use calibrated spectra 
to estimate the bolometric flux, which in turn allows us to determine 
the effective temperature nearly independently of atmospheric models.
In our companion paper on the detection of 
the large frequency separation \citep{bruntt08a}, we apply the parameters 
found here in our theoretical modelling of the star.

The presence of strong magnetic fields in roAp stars is thought to suppress
the turbulence in their outer atmospheres \citep{michaud70}. 
For this reason, enhanced diffusion leads to stratification of elements (\eg\ \citealt{babel92}).
A more complicated picture was found in the roAp star HR~3831 by \cite{koch04},
who monitored significant changes in spectral lines with rotational phase.
Using a Doppler imaging technique they found
rings and spots of different spatial scales, 
with enhanced abundances up to 7 dex for some elements (see also \citealt{lueft07}).
Such studies require high-quality data covering a wide 
range of rotational phases and this has not been done for \acir. 
Single spectra indicate that it is a typical roAp star, 
with high abundances of certain elements such as Co and Nd \citep{kupka96}.
To confirm the evidence of spots on \acir\ found by \cite{koch01b},
we present in Sect.~\ref{sec:vwa} an analysis of two high-resolution and high-S/N spectra 
taken at different rotational phases.

%
%

\section{Observations and Data Reduction}
\label{sec:obs}

The Sydney University Stellar Interferometer (SUSI; \citealt{davis99}) was 
used to measure the squared visibility,
\ie\ the normalised squared modulus of the complex visibility or $V^2$, 
on a total of eight nights. 
The red-table beam-combination system was employed with a filter of centre 
wavelength and full-width half-maximum 700\,nm and 80\,nm, respectively. 
This system, including the standard SUSI observing procedure, data reduction
and calibration, is described by \cite{davis07}. 


Target observations were bracketed with calibration measurements of 
nearby stars, which were chosen to be essentially unresolved.
Using an intrinsic colour interpolation (and spread in data) of 
measurements made with the Narrabri Stellar Intensity Interferometer 
\citep{hanbury74}, the angular diameters (and associated uncertainty) of the 
calibrator stars were estimated and corrected for the effects of 
limb-darkening. However, post-processing indicated that two of the 
calibrator stars (HR~4773: $\gamma$~Mus and HR~5132: $\epsilon$~Cen)
are binaries and could not provide adequate calibration information.
The adopted stellar parameters of the remaining calibrator stars are given in Table~\ref{cal_table}. 
A linear interpolation of the calibrator transfer functions 
was then used to scale the observed squared-visibility of the target.
This procedure resulted in a total 
of 60 estimations of $V^2$ and a summary of each night is given in 
Table~\ref{tab:obs}.

This calibration is not ideal due to (i) the latency between calibration
and target measurements; and (ii) the use of only one calibrator on all but two nights. 
However, the use of a nonlinear function to correct partially for residual 
seeing effects \citep{ireland06} as part of the data reduction reduces 
any systematic error introduced. 

It should also be noted that the 
uncertainty in the calibrator angular diameters 
only has a small effect on the final, calibrated target $V^2$.
Since our calibrators are nearly unresolved, their true visibilities 
are close to unity and depend only weakly on their actual diameters.
Therefore, propagation of these uncertainties is greatly reduced 
and only has a minor effect on the final uncertainty of the diameter of \acir.

\begin{table}
\centering
    \caption{Adopted parameters of the two calibrator stars.
The primary calibrator, \betacir, was used on all nights 
while the secondary calibrator, \sigmalup, was also used on 22 April and 30 July.} 
    \label{cal_table}
    \begin{tabular}{@{}cccccc@{}}
	\hline
    HR 	 & Name 	& Spectral & $V$    & UD Diameter       & Separation\\
	 &		&   Type   &(mag)   &  (mas)    	& from $\alpha$ Cir\\
	\hline
    5670 & \multicolumn{1}{r}{\betacir}  & A3\,V      & 4.05 & $0.58 \pm 0.05$ &  \multicolumn{1}{r}{ 7\fdg42} \\
    5425 & \multicolumn{1}{r}{\sigmalup} & B2\,III    & 4.42 & $0.20 \pm 0.04$ &  \multicolumn{1}{r}{14\fdg59} \\
	\hline
    \end{tabular}
\end{table}

\section{Angular Diameter}
\label{sec:fit}


In the simplest approximation, the brightness distribution of a star can be modelled
as a disc of uniform (UD) irradiance with angular diameter $\theta_{\rm UD}$.
A two-aperture interferometer has a theoretical response to such a 
model given by
\begin{equation}
\label{udisc_v2}
|V|^2 = \left |A\frac{2 J_1(\pi |\bmath{b}| \theta / \lambda)}
            {\pi |\bmath{b}| \theta / \lambda} \right|^2,
\end{equation}
where $\lambda$ is the observing wavelength, $\bmath{b}$ is the baseline
vector projected onto the plane of the sky and $J_1$ is a first-order Bessel function. 
For stars with a compact atmosphere, only small corrections 
are needed to account for monochromatic limb-darkening (see \citealt{davis00}). 
The parameter $A$ is included in the model to account for 
(i) any incoherent flux from the nearby faint
K5\,V star \gjstar\ that could enter the interferometer's 
field of view\footnote{\cite{poveda94} list the magnitude $V=8.47$, 
while \cite{gould04} list $V=9.46$ and a separation from \acir\ of $15.7$''.}; and
(ii) any instrumental effects arising from the differing spectral types of \acir\ and the calibrator stars.
The latter effect was found to be negligible as the spectral types are very similar (see also \citealt{davis07}).


Eq.~\ref{udisc_v2} is strictly valid only for monochromatic observations,
but is an excellent approximation when the effective wavelength is 
correctly defined (see \citealt{tango02, davis07}). 
During the \acir\ observations, 
the coherent field-of-view of SUSI was found to be greater than 
6.6\,mas, which is larger than the angular extent of \acir\ 
(see below). Hence, bandwidth smearing can be considered negligible. 
The interferometer's effective wavelength when observing an 
A-type main-sequence star is approximately $696.0\pm 2.0$\,nm \citep{davis07}.

\begin{table}
  \centering
    \caption{Summary of observational interferometry data for \acir\ from SUSI. 
             The night of the observation is 
             given in Columns 1 and 2 as a calendar date and the mean MJD$-54\,000$. 
	     The nominal and mean projected baselines in units of metres are
	     given in Columns 3 and 4, respectively. The weighted-mean 
	     squared visibility, associated error and the number of observations 
	     during each night are given in the last three columns.}
    \label{tab:obs}
    \begin{tabular}{@{}llcrccr}
	\hline
Date  	 &      & Nominal  & \multicolumn{1}{c}{Projected}  &              &                 &                             \\
in 2007  & MJD	& Baseline & \multicolumn{1}{c}{Baseline}   & ${\bar V^2}$ & ${\bar \sigma}$ & \multicolumn{1}{c}{$N$}\\	
	\hline
April~18  & $208.62$ & \multicolumn{1}{r}{$ 5$} & ${ 4.08}$ &  $0.964$ & $0.012$ & $10$ \\
April~20  & $210.63$ & \multicolumn{1}{r}{$20$} & ${16.37}$ &  $1.039$ & $0.016$ & $ 9$ \\
April~22  & $212.66$ & \multicolumn{1}{r}{$20$} & ${16.29}$ &  $0.900$ & $0.015$ & $ 9$ \\
May~24    & $244.56$ & \multicolumn{1}{r}{$80$} & ${65.25}$ &  $0.533$ & $0.008$ & $ 8$ \\
May~25    & $245.55$ & \multicolumn{1}{r}{$80$} & ${65.34}$ &  $0.528$ & $0.010$ & $ 8$ \\
June~22   & $273.46$ & \multicolumn{1}{r}{$ 5$} & ${ 4.10}$ &  $0.987$ & $0.024$ & $ 4$ \\
July~30   & $311.40$ & \multicolumn{1}{r}{$40$} & ${32.48}$ &  $0.845$ & $0.038$ & $ 6$ \\
August~4  & $316.40$ & \multicolumn{1}{r}{$80$} & ${64.42}$ &  $0.560$ & $0.013$ & $ 6$ \\
     \hline
    \end{tabular}
\end{table}

\begin{figure}
\includegraphics[width=\linewidth]{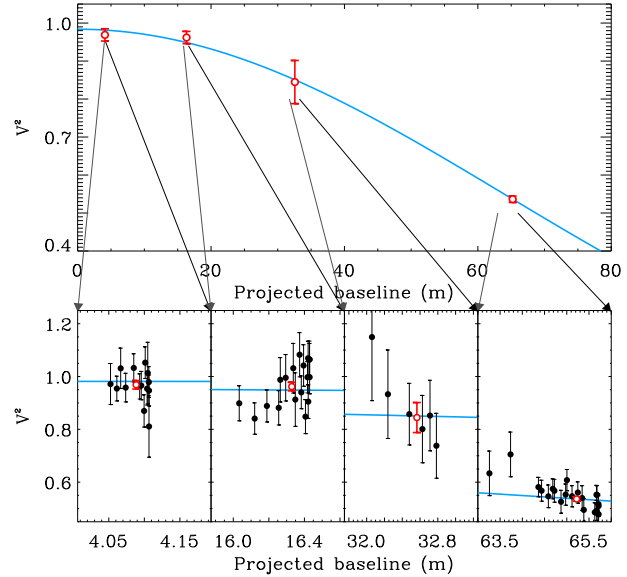} 
\caption{\label{fig.vis} The top panel shows the baseline-averaged $\bar V^2$ measures (open circles) 
with the fitted uniform disc model overlaid (solid line). 
The bottom panels show the individual $V^2$ measures (filled circles) at each baseline.}
\end{figure}

We estimated $\theta_{\rm UD}$ and $A$ by
fitting Eq.~\ref{udisc_v2} to the 60 individual measures of $V^2$.
The fit was achieved using a $\chi^2$ minimisation 
with an implementation of the Levenberg-Marquardt method. 
The reduced $\chi^2$ of the fit was $2.2$, 
implying that the measurement uncertainties were underestimated. 
We have therefore multiplied the $V^2$ measurement uncertainties
by $f_\chi=\sqrt{2.2}$ to obtain a reduced $\chi^2$ of unity.
In Fig.~\ref{fig.vis} the fitted uniform disc model (solid line)
is compared to the weighted-mean $\bar V^2$ measures at each of the four baselines.
The bottom panels show the individual measures at each baseline. 
The plotted error bars are scaled by $f_\chi$.

Formal uncertainties in $\theta_{\rm UD}$ and $A$,
derived from the diagonal elements of the covariance matrix,
may be underestimates for three reasons: 
the visibility measurement errors may not strictly conform to a normal distribution, 
Eq.~\ref{udisc_v2} is non-linear, 
and the primary calibrator is partially resolved.
The effect of these factors on the model parameter uncertainties was 
investigated using Markov chain Monte Carlo (\mcmc) simulations. 
This method involves a likelihood-based random walk through parameter space,
using an implementation of a Metropolis-Hastings algorithm,
and yields the full marginal posterior probability density function 
(PDF; for an introduction to \mcmc\ see Chapter 12 of \citealt{Gregory05a}).
Furthermore, the current knowledge 
of the system can be included in the analysis by assuming an a~priori distribution. 
For example, knowledge of the effective wavelength and 
calibrator angular diameter (and associated uncertainties) can be included 
into the uncertainty estimation of the model parameters. 

Over 25 \mcmc\ simulations were completed, each with $10^6$ iterations, and adopting 
Gaussian likelihood distributions for the effective wavelength and calibrator angular diameters.
The resulting PDFs for $\theta_{\rm UD}$ produced a 1$\sigma$ 
parameter uncertainty approximately twice as large as the formal uncertainty
derived from the covariance matrix, while those for $A$ were approximately equal.
Since the simulations take into account the uncertainties in the
effective wavelength and calibrator angular diameter, we adopt the standard
deviations from the \mcmc\ simulations. We believe they represent the
most realistic and conservative parameter uncertainty estimates for our dataset.

Final values of $\theta_{\rm UD}$ and $A$ are $1.063\pm0.034$\,mas and $0.992\pm0.006$, 
respectively. 
The value of $A$, while close to unity, indicates that the measurements 
are affected slightly by the incoherent flux of the faint star \gjstar.

The work of \cite{davis00} can be used to correct the angular diameter for
the effect of limb-darkening, provided the effective temperature (\teff), 
surface gravity (\logg) and metallicity (\feh) are known. 
Since roAp stars have non-standard atmospheres
and \cite{davis00} used \atlasni\ atmospheric models, we investigated the
limb-darkening correction within the following large region of
parameter space: \teff\,$=7250$ to $8500$\,K, \logg\,$= 4.0$ to $4.5$, 
\feh\, = $-0.3$ to $+0.3$. The limb-darkening correction of \cite{davis00},
at an observing wavelength 696\,nm, varies between $1.0344$ and $1.0426$
and is more sensitive to changes in \teff\ than in \logg\ or \feh. 
We conservatively set an uncertainty of $0.010$ in the limb-darkening 
correction, but note that this uncertainty is small compared to the
uniform-disc angular diameter uncertainty. 

For the fundamental parameters found in Sect.~\ref{sec:disc},
we get a limb-darkening correction of $1.039\pm0.010$, to obtain
the limb-darkened (LD) angular diameter of $\theta_{\rm LD} = 1.105\pm0.037$\,mas.

\section{The bolometric flux of \acir}

We derived the bolometric flux of \acir\ using two different datasets.
The first value was obtained by combining the observed ultraviolet flux
retrieved from IUE Newly Extracted Spectra (INES) data archive, with the theoretical
flux obtained from a Kurucz model (with IDL routine KURGET1). 
To compute the integrated fluxes, we followed the same method as \cite{north81}. 
The IUE measurements were used to compute the flux in the wavelength 
interval $1150 \,\mathring{\rm A} <\lambda < 3207 \,\mathring{\rm A}$, 
while the flux from $\lambda=3210 \,\mathring{\rm A}$ to infinity was computed 
from the Kurucz model that best fitted the seven intrinsic colours of the star in the 
Geneva photometric system, retrieved from the catalogue of \cite{rufener89}. 
Moreover, the energy distribution was extrapolated to the interval 
$912 \,\mathring{\rm A} <\lambda < 1150 \,\mathring{\rm A}$, assuming zero flux 
at $\lambda=912 \,\mathring{\rm A}$. 
The resulting integrated flux was 
$f_{\rm bol}=(1.28\pm 0.02)\times10^{-6}$erg~cm~$^{-2}$~s$^{-1}$, 
where the quoted uncertainty range corresponds to the difference 
between the maximum and minimum integrated fluxes allowed 
by the uncertainties in the parameters of the Kurucz model adopted, 
which, in turn, reflect the uncertainties in the photometric data.
 
The abnormal flux
distributions that are characteristic of Ap stars make determinations such as that described above, 
based on atmospheric models appropriate to normal stars, rather unreliable. When spectra calibrated in
flux are also available for visible wavelengths, a more reliable determination of the bolometric flux 
can in principle be obtained. Unfortunately, even though two low-resolution spectra calibrated in flux 
are available in the literature  for \acir\ \cite[see catalogues by][]{aleks96,burnashev85}, 
the errors associated with the calibrations in flux as function of wavelength are not given 
in the corresponding bibliographic sources. Nevertheless, we have calculated a second
value for the bolometric flux using the same method as above, 
but replacing the synthetic spectra obtained for the Kurucz model by the two
low-resolution spectra of \acir, calibrated in flux, retrieved from
the catalogues mentioned above. In this case, the flux as function of wavelength was obtained by 
combining the IUE measurements in the wavelength interval 
$1150 \,\mathring{\rm A} <\lambda < 3349 \,\mathring{\rm A}$, 
the mean of the two low-resolution spectra calibrated in flux in the interval 
$3200 \,\mathring{\rm A} <\lambda < 7350 \,\mathring{\rm A}$, and the flux derived from 
the mean of the two Kurucz models that best fitted each of the two low-resolution spectra, 
for wavelengths longer than $7370 \,\mathring{\rm A}$.
The integrated flux obtained in this way was
$f_{\rm bol}=(1.18\pm 0.01)\times10^{-6}$erg~cm~$^{-2}$~s$^{-1}$, where the uncertainties reflect 
the difference between the integrated fluxes derived when each of the two low-resolution spectra 
(and corresponding best-fit Kurucz model at higher wavelengths) were considered separately. 

Since the details and associated uncertainties of the calibrations in flux of the low-resolution spectra 
used in the latter calculation are unknown, we will take a conservative approach and allow 
the bolometric flux to vary within the two extremes of the values derived through 
the two different approaches. Hence, in what follows, we use for the bolometric flux, 
the mean value $f_{\rm bol}=(1.23\pm 0.07)\times10^{-6}$erg~cm~$^{-2}$~s$^{-1}$.

\begin{table}
    \caption{Physical parameters of \acir\label{tab:param}. All parameters are from
the current study except the \hipp\ parallax \citep{leeuwen07}.}
\centering
    \begin{tabular}{@{}lr@{$\pm$}lcl@{}}
	\hline
	Parameter & \multicolumn{2}{c}{Value}  & Uncertainty (\%) \\
	\hline
	$\theta_{\rm LD}$ (mas)   &  $1.105$ & $0.037$	   & 3.4 \\ 
	$f_{\rm bol}$ ($10^{-9}$\,Wm$^{-2}$)&  $1.23$  & $0.07$	   & 5.7 \\ 
	$\pi_{\rm p}$ (mas) 	  & $60.36$  & $0.14$      & 0.2 \\ 
\hline
	\teff\ (K)                & $7420$   & $170$       & 2.3 \\ 
	$R$ (R$_{\odot}$)         & $1.967$  & $0.066$     & 3.4 \\ 
	$L$ (L$_{\odot}$)         & $10.51$  & $0.60$	   & 5.7 \\ 
	$M$ (M$_{\odot}$)         & $1.7$    & $0.2$       & 12  \\ 
	\hline


    \end{tabular}
\end{table}

\begin{figure}
\includegraphics[width=\linewidth]{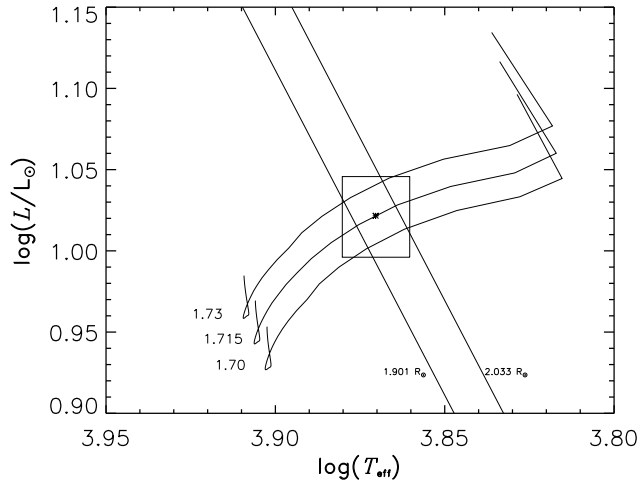} 
\caption{\label{HR} 
The position of \acir\ in the Hertzsprung-Russell diagram, 
with three evolution tracks for models with masses of 1.70, 1.715 and 1.73 M$_\odot$ for comparison. The constraints on the
fundamental parameters are indicated by the 
$1\sigma$-error box (\teff, $L/L_\odot$) and the diagonal lines (radius).}
\end{figure}


\begin{table*}
    \caption{Fundamental atmospheric parameters of \acir\ as found
in the literature and the current study.\label{tab:fund}}
    \begin{tabular}{@{}lr@{}l|l@{}l|r@{}ll|l}
	\hline
Study &  \multicolumn{2}{c}{\teff\ [K]} &  \multicolumn{2}{c}{\logg} & \multicolumn{2}{c}{\feh} & \multicolumn{1}{l}{Applied method} \\
	\hline
\cite{kurtz93}     &   $8000$ & $      $  &        &           & $  $&$   $          & \hbeta\ index     \\ 
\cite{north94}     &   $8000$ & $      $  & $4.42$ & $       $ & $+0$&$.2 $          & Geneva indices    \\ 
\cite{kupka96}     &   $7900$ & $\pm200$  & $4.2 $ & $\pm0.15$ & $-0$&$.13\pm0.10$   & Spectral analysis \\
$-$                &   $7950$ &           & $4.3 $ & $       $ & $  $&$   $          & \str\ indices     \\
$-$                &   $7880$ &           & $4.4 $ & $       $ & $+0$&$.3 $          & Geneva indices    \\ 
\cite{soko98}      &   $8440$ & $\pm290$  & $    $ & $       $ & $  $&$   $          & Continuum slope   \\ 
$-$                &   $7680$ & $      $  & $    $ & $       $ & $  $&$   $          & Geneva indices    \\ 
\cite{matt99}      &   $8000$ & $\pm100$  & $    $ & $       $ & $  $&$   $          & \hbeta\ index     \\
\cite{kochukhov06} &   $7673$ & $\pm200$  & $4.12$ & $\pm0.06$ & $  $&$   $          & Geneva indices    \\ 
This study         &   $7420$ & $\pm170$  & $4.09$ & $\pm0.08$ & $  $&$   $          & Interferometry $+$ bolometric flux\\
$-$                &          &           &        &           & $-0$&$.12\pm0.08$   &  Spectral analysis    \\
\hline

\end{tabular}
\end{table*}

%
%

\section{Fundamental parameters of \acir}
\label{sec:disc}

We use the measured angular diameter and bolometric flux to estimate \teff\ 
from the definition
\begin{equation}
\sigma T_{\rm eff}^4 = 4 f_{\rm bol} \, / \, \theta_{\rm LD},
\end{equation}
where $\sigma$ is the Stefan-Boltzmann constant. 
We can also obtain estimates of the radius and luminosity from the measured 
angular diameter and bolometric flux, when they are combined with the 
parallax:
\begin{equation}
R = 2\theta \, {C} / {\pi_{\rm p}},
\end{equation}
and
\begin{equation}
L = 4 \pi \, f_{\rm bol} \, {C^2} / {\pi_{\rm p}^2},
\end{equation}
where $C$ is the conversion from parsecs to metres. 
We have used a parallax of $\pi_{\rm p} = 60.36\pm0.14$~mas, 
based on the new reduction of the raw  \hipp\ data \citep{leeuwen07}.
It has lower uncertainty, but is in good
agreement with the original data release \citep{ESA97}, 
which was $\pi_{\rm p} = 60.97\pm0.58$~mas.
The values we have determined for the fundamental parameters 
of \acir, along with the observable quantities, are given in Table~\ref{tab:param}.
Due to the accurate parallax, the uncertainties in the radius and luminosity 
are dominated by the uncertainty in the angular diameter and bolometric flux, respectively.
The uncertainty in the effective temperature is
determined by the uncertainties in both the angular diameter and the bolometric flux.


In Fig.~\ref{HR} we show the location of \acir\ in the Hertzsprung-Russell diagram. 
The 1$\sigma$-error box and diagonal lines are based on the fundamental parameters 
and associated errors derived in this study (see Table~\ref{tab:param}).
Three \cesam\ \citep{morel97} evolutionary tracks crossing the error box for \acir\ are also shown.
The tracks were calculated for solar metallicity using a mixing 
length parameter of $\alpha = 1.6$.  Diffusion and rotation were not included. 
Note that changes in the model input parameters can significantly modify 
the evolutionary tracks for models with the same mass. In a study of the 
roAp star HR~1217, \cite{cunha03} found that, while changes in the
mixing length parameter and convective overshoot had a minor impact
on the mass derived from model fitting of the classical observables,
the uncertainties in the initial helium abundance and in the global 
metallicity of the star resulted in a significant uncertainty in the 
mass determination that in the worst case could be as large as 10\%. 
Therefore, from Fig.~\ref{HR} and the results of \cite{cunha03}, we 
conservatively estimate the mass of $\alpha$ Cir to be 
$M = 1.7\pm 0.2$\,M$_\odot$. 
Even so, we obtain an accurate surface gravity of $\log g = 4.09\pm0.08$.
This is a consequence of the accurately measured interferometric radius.


\subsection{Comparison with previous estimates}

Previous studies of \acir\ have used either spectroscopy or photometric indices to
estimate the fundamental atmospheric parameters. 
We list some recent estimates from the literature in Table~\ref{tab:fund}.
\cite{kurtz93} and \cite{matt99} used the \str\ \hbeta\ index to estimate \teff.
\cite{north94} made an abundance analysis of \acir\ based on a few selected
wavelength regions with high resolution, but the fundamental parameters were 
fixed based on Geneva photometry. 
\cite{kupka96} analysed high-resolution spectra of \acir\
and estimated the fundamental parameters using Geneva and \str\
photometric indices, the Balmer lines and a detailed classical analysis of several iron-peak lines.
\cite{soko98} proposed a new method to estimate \teff\ of roAp stars based on 
the slope of the continuum near the Balmer jump and found a high \teff\ for \acir.
\cite{kochukhov06} analysed a large sample of magnetic chemically peculiar stars
and they used Geneva indices to estimate the fundamental parameters of \acir.
Some of the studies mentioned here that used photometric indices 
have not given uncertainties on the parameters.
Realistic values are about 200~K, $0.2$~dex, and $0.1$~dex on \teff, \logg, and \feh, 
respectively \citep{kupka01}. \cite{koch05} explored the influence of magnetic fields
on the photometric indices and found relatively small corrections, 
especially for field strengths of a few kG or less, as found in \acir\ (see Sect.~\ref{sec:mag}).

Previous estimates of \logg\ lie in the range from $4.1$ to $4.6$, 
considering 1$\sigma$ error bars. We have obtained an accurate 
estimate, \logg\,$=4.09\pm0.08$, which is lower than previous
estimates but in general agreement.
The range in \teff\ estimated for \acir\ in the literature is $7470$ to $8730$\,K, 
considering 1$\sigma$ error bars.
This has hampered detailed studies of \acir\ that rely on theoretical models, 
such as spectroscopic analysis (atmospheric models) or asteroseismology (evolution and pulsation models).
In general the Geneva colours give lower \teff\ than the \str\ or \hbeta\ indices.
The slight differences in the parameters found using the same photometric system are due to different calibrations being used. 
Our value, \teff\,$=7420\pm170$\,K, is the lowest of all temperature determinations. 
However, it is in acceptable agreement with the \teff\ from 
Geneva indices (the mean value from four estimates is $7810\pm160$\,K),
and barely agrees with \teff\,$=7900\pm200$\,K found by \cite{kupka96} based on
spectroscopic analysis of Fe\iet\ and Fe\ito\ lines.
We note that a higher \teff\ is also implied in our analysis (in Sect.~\ref{sec:vwa}) 
of Fe\iet\ spectral lines  due to a strong correlation with excitation potential. 
This could either imply stratification of Fe, as seen in other roAp stars, or, 
alternatively, that \teff\ of the atmospheric model is too low by $600$~K.
We believe the higher \teff\ found by \cite{kupka96} is explained by this.

We stress that our estimates of \logg\ and \teff\ are 
largely model-independent, unlike the previous estimates. 
Therefore we recommend that these values be used in future 
studies of \acir. Our value for \teff\ depends on the
angular diameter from interferometry and the bolometric flux. 
The latter estimate should be confirmed by making new observations.

   \begin{table}
      \caption[]{The atomic number, element name, 
wavelength, and oscillator strength (\loggf) from the \vald\ database for lines used in the abundance analysis.
{\em The full version of this Table is available in the on-line version.}
         \label{tab:lines}}
\centering                          
\begin{tiny}
\begin{tabular}{r@{\hskip 0.23cm}c@{\hskip 0.23cm}c|r@{\hskip 0.23cm}c@{\hskip 0.23cm}c|r@{\hskip 0.23cm}c@{\hskip 0.23cm}c}
\hline\hline

\eeee & $\lambda$ [\AA] & \loggf & \eeee & $\lambda$ [\AA] & \loggf & \eeee & $\lambda$ [\AA] & \loggf \\

\hline
 $^{ 6}$C\ione & 5017.090 & $-2.500$ & $$     Ti\itwo & 5490.690 & $-2.650$ & $$     Fe\ione & 5217.389 & $-1.070$  \\
 $$     C\ione & 5540.751 & $-2.376$ &  $^{23}$V\ione & 5670.853 & $-0.420$ & $$     Fe\ione & 5232.940 & $-0.058$  \\
 $$     C\ione & 5603.724 & $-2.418$ &  $$     V\ione & 6081.441 & $-0.579$ & $$     Fe\ione & 5242.491 & $-0.967$  \\
 $$     C\ione & 5643.368 & $-2.670$ &  $$     V\ione & 6199.197 & $-1.300$ & $$     Fe\ione & 5250.646 & $-2.181$  \\
 $$     C\ione & 5800.602 & $-2.338$ & $^{24}$Cr\ione & 5204.506 & $-0.208$ & $$     Fe\ione & 5253.462 & $-1.573$  \\
 $$     C\ione & 6413.547 & $-2.001$ & $$     Cr\ione & 5224.972 & $-0.096$ & $$     Fe\ione & 5263.306 & $-0.879$  \\
 $^{ 8}$O\ione & 5330.737 & $-0.984$ & $$     Cr\ione & 5247.566 & $-1.640$ & $$     Fe\ione & 5269.537 & $-1.321$  \\
 $$     O\ione & 6155.986 & $-1.120$ & $$     Cr\ione & 5296.691 & $-1.400$ & $$     Fe\ione & 5281.790 & $-0.834$  \\
 $$     O\ione & 6156.776 & $-0.694$ & $$     Cr\ione & 5297.376 & $+ 0.167$ & $$     Fe\ione & 5283.621 & $-0.432$ \\
 $$     O\ione & 6158.186 & $-0.409$ & $$     Cr\ione & 5348.312 & $-1.290$ & $$     Fe\ione & 5288.525 & $-1.508$  \\
$^{11}$Na\ione & 5682.633 & $-0.700$ & $$     Cr\ione & 5787.965 & $-0.083$ & $$     Fe\ione & 5302.302 & $-0.720$  \\
$$     Na\ione & 5688.205 & $-0.450$ & $$     Cr\ione & 6661.078 & $-0.190$ & $$     Fe\ione & 5315.070 & $-1.550$  \\
$^{12}$Mg\ione & 5172.684 & $-0.402$ & $$     Cr\itwo & 5237.329 & $-1.160$ & $$     Fe\ione & 5324.179 & $-0.103$  \\
$$     Mg\ione & 5183.604 & $-0.180$ & $$     Cr\itwo & 5246.768 & $-2.466$ & $$     Fe\ione & 5353.374 & $-0.840$  \\
$$     Mg\ione & 5528.405 & $-0.620$ & $$     Cr\itwo & 5305.853 & $-2.357$ & $$     Fe\ione & 5364.871 & $+ 0.228$ \\
$$     Mg\ione & 5711.088 & $-1.833$ & $$     Cr\itwo & 5308.408 & $-1.846$ & $$     Fe\ione & 5365.399 & $-1.020$  \\
$^{13}$Al\ione & 5557.063 & $-2.110$ & $$     Cr\itwo & 5310.687 & $-2.280$ & $$     Fe\ione & 5367.467 & $+ 0.443$ \\
$$     Al\ione & 6696.023 & $-1.347$ & $$     Cr\itwo & 5313.563 & $-1.650$ & $$     Fe\ione & 5373.709 & $-0.860$  \\
$^{14}$Si\ione & 5665.555 & $-2.040$ & $$     Cr\itwo & 5334.869 & $-1.562$ & $$     Fe\ione & 5383.369 & $+ 0.645$ \\
$$     Si\ione & 5708.400 & $-1.470$ & $$     Cr\itwo & 5407.604 & $-2.151$ & $$     Fe\ione & 5386.334 & $-1.770$  \\
$$     Si\ione & 5772.146 & $-1.750$ & $$     Cr\itwo & 5420.922 & $-2.458$ & $$     Fe\ione & 5389.479 & $-0.410$  \\
$$     Si\ione & 6125.021 & $-0.930$ & $$     Cr\itwo & 5478.365 & $-1.908$ & $$     Fe\ione & 5391.461 & $-0.825$  \\
$$     Si\ione & 6142.483 & $-0.920$ & $$     Cr\itwo & 5508.606 & $-2.110$ & $$     Fe\ione & 5393.168 & $-0.715$  \\
$$     Si\ione & 6414.980 & $-1.100$ & $$     Cr\itwo & 5510.702 & $-2.452$ & $$     Fe\ione & 5398.279 & $-0.670$  \\
$$     Si\ione & 6856.037 & $-1.060$ & $$     Cr\itwo & 6068.023 & $-1.736$ & $$     Fe\ione & 5400.502 & $-0.160$  \\
 $^{16}$S\ione & 6395.163 & $-2.040$ & $$     Cr\itwo & 6070.105 & $-2.944$ & $$     Fe\ione & 5405.775 & $-1.844$  \\
 $$     S\ione & 6396.644 & $-1.820$ & $^{25}$Mn\ione & 5377.637 & $-0.109$ & $$     Fe\ione & 5410.910 & $+ 0.398$ \\
 $$     S\ione & 6415.522 & $-1.360$ & $$     Mn\ione & 5413.689 & $-0.589$ & $$     Fe\ione & 5415.199 & $+ 0.642$ \\
 $$     S\ione & 6743.531 & $-0.920$ & $$     Mn\ione & 5420.355 & $-1.462$ & $$     Fe\ione & 5424.068 & $+ 0.520$ \\
 $$     S\ione & 6757.171 & $-0.310$ & $$     Mn\ione & 5470.637 & $-1.702$ & $$     Fe\ione & 5434.524 & $-2.122$  \\
$^{20}$Ca\ione & 5512.980 & $-0.712$ & $$     Mn\ione & 5537.760 & $-2.017$ & $$     Fe\ione & 5466.396 & $-0.630$  \\

\hline
\end{tabular}
\end{tiny}
\end{table}

%
%


\section{Spectral analysis of \acir}
\label{sec:vwa}

We have obtained high quality spectra of \acir\ with the 
Ultraviolet and Visual Echelle Spectrograph (UVES) on the 
Very Large Telescope (VLT). 
The spectra were collected as part of asteroseismic campaigns in 2004 \citep{kurtz06newtype} and 2006. 
In the latter case, the spectra were used to compare the variability of
\acir\ in radial velocity and photometry from the
star tracker on the \wire\ satellite and from the ground \citep{bruntt08a}.
The spectrum from 2004 is comprised of $64$ co-added spectra collected over $0.5$\,h, 
at a mean rotational phase of $0.230$. We define phase zero to be at minimum light 
in the rotational modulation curve measured with \wire\ \citep{bruntt08a}.
The spectrum from 2006 is comprised of $149$ co-added spectra collected over $2.4$\,h at mean rotational phase $0.975$,
\ie\ close to minimum light.
The resulting spectra have $S/N$ of $800$ and $1100$, respectively, in the continuum around 5500\,\AA.
The spectral resolution is about $R=110\,000$ and we used 
spectra from the red arm in the wavelength ranges $4970 - 5950$\,\AA\ and $6090 -6900$\,\AA. 
We avoided the range $5880 - 5950$\,\AA, which is affected by many telluric lines.
This was especially the case for the 2006 spectrum, which was collected through cloud.
For this reason, we base our final results on the spectrum from 2004.
The spectra were carefully normalized by identifying continuum windows 
in a synthetic spectrum with the same parameters as \acir.

The spectral lines were analysed using the \vwa\ package \citep{bruntt04, bruntt07}.
The abundance analysis relies on the calculation of synthetic spectra
with the \synth\ code by \cite{valenti96}. 
Atomic parameters and line broadening coefficients were extracted from the \vald\ database \citep{vald}. 
Table~\ref{tab:lines} lists the lines we have used in the analysis.
We used modified \atlasni\ models from \cite{heiter02} with the adopted 
parameters from interferometry \teff\,$=7420\pm170$\,K and \logg\,$=4.09\pm0.08$. 
We initially assumed solar metallicity, but adjusted the abundances of individual elements 
in the model after a few iterations. We adjusted the microturbulence to $\xi_t=1.60\pm0.15$\,\kms\
by requiring that there is no correlation between the abundance and strength 
of 50 weak Fe\iet\ lines with equivalent widths $< 90$\,m\AA\ and 
excitation potentials in the range 3--5\,eV.
We repeated our analy\-sis with \vwa\ using the \llmodels\ code \citep{shulyak04},
and found the same result for all elements, with differences below 0.03~dex.

\begin{figure*}
 \includegraphics[width=16cm]{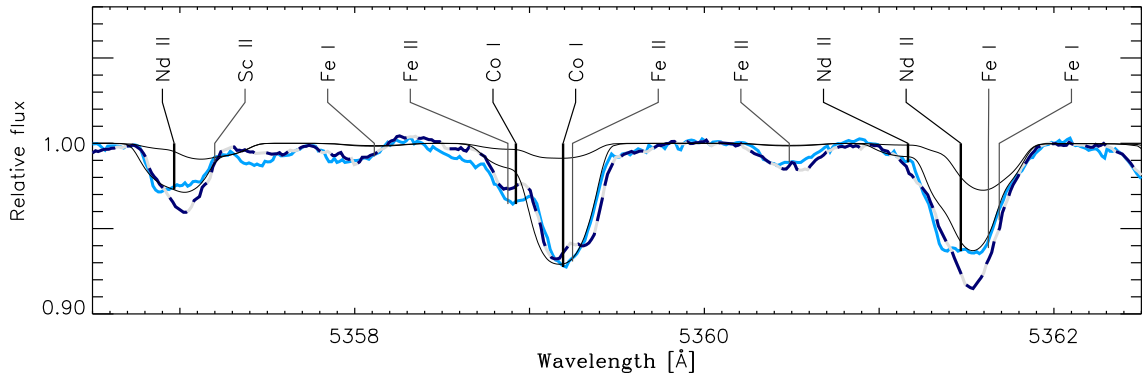}  
\caption{\label{fig:vwa} 
The two observed UVES spectra of \acir\ are compared (solid and dashed thick lines) in a region with 
strong neodymium and cobalt lines. Also shown are two synthetic spectra (thin lines) 
for solar abundance and when the abundance of Nd and Co is increased 
by about a factor of 10.}
\end{figure*}

\begin{table}
    \caption{Abundance of \acir\ relative to the Sun based on the analysis of the spectrum from 2004.
The mean abundance and the number of neutral and ionized lines are denoted I and II.
\label{tab:vwa}}
\centering
    \begin{tabular}{llr|lr}
	\hline
Element & \multicolumn{1}{c}{[A]$_{\rm I}$} & $N_{\rm I}$ & \multicolumn{1}{c}{[A]$_{\rm II}$} & $N_{\rm II}$ \\
	\hline

 C & $ +0.05\pm0.07$ & $   6$ & $             $ & $    $ \\
 O & $ -0.15\pm0.09$ & $   4$ & $             $ & $    $ \\
 Na& $ -0.21\pm0.14$ & $   2$ & $             $ & $    $ \\
 Mg& $ -0.01\pm0.17$ & $   2$ & $             $ & $    $ \\
 Al& $ +0.18\pm0.19$ & $   2$ & $             $ & $    $ \\
 Si& $ -0.14\pm0.20$ & $   7$ & $             $ & $    $ \\
 S & $ +0.35\pm0.08$ & $   5$ & $             $ & $    $ \\
 Ca& $ +0.19\pm0.10$ & $  20$ & $             $ & $    $ \\
 Sc& $             $ & $    $ & $ -0.42\pm0.11$ & $   3$ \\
 Ti& $ -0.20\pm0.13$ & $   5$ & $ -0.17\pm0.09$ & $   5$ \\
 V & $ +0.60\pm0.21$ & $   3$ & $             $ & $    $ \\
 Cr& $ +0.48\pm0.11$ & $   8$ & $ +0.49\pm0.10$ & $  11$ \\
 Mn& $ +0.68\pm0.14$ & $   7$ & $ +0.68\pm0.09$ & $   4$ \\
 Fe& $ -0.12\pm0.08$ & $ 125$ & $ -0.03\pm0.05$ & $  20$ \\
 Co& $ +1.20\pm0.08$ & $  21$ & $             $ & $    $ \\
 Ni& $ -0.58\pm0.10$ & $   7$ & $             $ & $    $ \\
 Cu& $ -1.02\pm0.18$ & $   2$ & $             $ & $    $ \\
 Y & $             $ & $    $ & $ +0.88\pm0.12$ & $  10$ \\
 Zr& $             $ & $    $ & $ +0.37\pm0.23$ & $   1$ \\
 Ba& $             $ & $    $ & $ -0.70\pm0.18$ & $   2$ \\
 Ce& $             $ & $    $ & $ +0.74\pm0.13$ & $   3$ \\
 Nd$^*$& $             $ & $    $ & $ +1.03\pm0.10$ & $  12$ \\
 Eu& $             $ & $    $ & $ +1.58\pm0.15$ & $   2$ \\ \hline
\multicolumn{5}{l}{$^*$From three Nd\itre\ lines we find [Nd]$=+2.06\pm0.12$.} \\

    \end{tabular}
\end{table}
%

The abundances relative to the Sun
for 23 species\footnote{We quote the abundance of an ion $A$ as $[A] = \log N_{\rm A}/N_{\rm tot} - (\log N_{\rm A}/N_{\rm tot})_\odot$, where $N$ is the number of atoms and the solar values are taken from \cite{grevesse07}.}
are given in Table~\ref{tab:vwa}.
To estimate uncertainties in the abundances we measured abundances for
model atmospheres with higher values of \teff\ ($+300$\,K), \logg\ ($+0.3$\,dex), and microturbulence ($+0.4$\,\kms),
and made a linear interpolation using the uncertainties of these parameters.
In addition, we added $0.17/\sqrt{N}$\,dex to the uncertainty, where $N$ is the number of lines,
and $0.17$ is the average \rms\ standard deviation on the abundances for four 
elements with the most spectral lines, \ie\ Ca, Fe, Co, and Nd.
For example, for Fe\ione\ the estimated uncertainty on the abundance is $0.08$\,dex,
including contributions from the uncertainty on the fundamental parameters.


The Fe\iet\ lines are the most numerous in the observed spectra.
We find a significantly lower abundance from Fe\iet\ lines 
with low excitation potential, $\epsilon$. For lines with 
$1<\epsilon\,[{\rm eV}]<3$ the abundance relative to the Sun is 
$-0.33\pm0.02$ (28 lines) and for $3<\epsilon\,[{\rm eV}]<5$ the abundance is 
$-0.07\pm0.02$ (89 lines).
The quoted uncertainties are the \rms\ on the mean value, and represent only the internal error.
The difference between low and high excitation lines is significant and 
could be an indication of vertical stratification, 
meaning the abundance of Fe is higher in the deeper layers, as discussed by \cite{ryab02, ryab07}.


In Fig.~\ref{fig:vwa} we show a small region of the observed spectra from 2004 (solid thick line) and 2006 (dashed thick line), 
with identification of the elements causing the lines. 
We show two synthetic spectra (thin lines) with different abundances of 
cobalt and the rare-earth element neodymium.
In the first synthetic spectrum the abundances are solar and in the
second we have adjusted the abundances of Co and Nd to fit the observed spectrum from 2004. 
We find over-abundances of Co\iet\ and Nd\itwo\ of $+1.20\pm0.08$ and $+1.03\pm0.11$ using 21 and 12 lines,
respectively. This is in acceptable agreement with the analysis by \cite{kupka96} who found 
$+1.62\pm0.20$ and $+1.24\pm0.25$, using 6 lines of each ion.
For other ions listed in Table~\ref{tab:vwa} we find general 
agreement with the previous detailed study by \cite{kupka96}.

The two spectra we analysed yield the same mean abundances, when averaging over several lines,
with the mean abundances differing by less than $0.05$--$0.10$\,dex in the two spectra. 
A difference of $0.20$\,dex was found for V\iet\ and Eu\itwo, 
but the analysis is based on just 3 and 2 lines, respectively.
It is interesting to note that the equivalent widths and detailed shapes of the Co\iet\ and Nd\itwo\ lines 
are significantly different in the two observed UVES spectra shown in Fig.~\ref{fig:vwa}. 
This indicates the presence of spots on the surface, as first reported by \cite{koch01b}.
A more detailed study of \acir\ is required to explore this, as was done for the roAp star HR~3831 by \cite{koch04}.

\subsection{Limits on the magnetic field strength\label{sec:mag}}

%

The roAp stars are magnetic, with surface field strengths as high as $24.5$\,kG in HD\,154708 \citep{hubrig05}, but more typically of several kG (see \citealt{kurtz06star154708}).  Knowledge of the magnetic field strength and structure is important for the study of $\alpha$\,Cir. A variable longitudinal magnetic field was discovered in this star by \cite{borra75} and \cite{wood75}. The latter paper shows a variable field with negative polarity and field strength extremum of $-1.5$\,kG, but with a large uncertainty. \cite{borra75} obtained more precise results using a photoelectric Balmer line polarimeter and could not confirm the extremum obtained by \cite{wood75}. They made seven measurements over seven consecutive nights, from which they detected a variable field ranging from $-560$ to $+420$\,G. From these observations they proposed a rotation period of the order of 12\,d -- a value we now know to be incorrect. \cite{mathys91} attempted to test the possibility of a short rotation period close to 1\,d by observing the longitudinal field strength. He obtained five spectra in three nights and did not detect any magnetic field within the errors of his observations. \cite{mathyshubrig97} presented two more longitudinal field measurements which show a negative polarity magnetic field, but their results were still below the 3$\sigma$ level.

From these observations it is clear that the star does have a magnetic field and that the longitudinal component is rather weak. The precision of the published observations is insufficient, however, to prove variability with the known rotation period of $P_{\rm rot}=4.46$\,d \citep{kurtz94,bruntt08a}. \cite{bychkov05}, for example, combined the data and derived a magnetic variation period equal to the known rotation period, but we conservatively are not convinced that rotational magnetic variability has been found, given the error bars of the individual measurements. While there is an indication of a rotational magnetic curve, additional high-precision observations of the longitudinal magnetic field for this bright star are sorely needed.

 
Estimating the magnetic field modulus $\langle H \rangle$ for $\alpha$\,Cir is also not a simple task, given its weak longitudinal field and rather high (for an roAp star) rotational velocity of $v \sin i = 13$\,\kms. No Zeeman components can be seen directly in our high resolution spectra, so we searched for the effect of magnetic broadening by measuring the full-width at half maximum (FWHM) for selected unblended spectral lines with a spread of Land\'e factors. This gives a value of the magnetic field similar to the mean magnetic field modulus obtained from split Zeeman components in slowly rotating stars where the components are resolved. From the measurement of the FWHM for 20 Fe and Cr lines we obtained a magnetic field modulus of $0.9\pm0.5$\,kG,
a significant improvement on the upper limit of $3$--$3.5$\,kG set by \cite{kupka96}, using a similar approach.

We estimated an upper limit for $\langle H \rangle$ by comparing average observed spectral lines with calculated lines in the presence of a magnetic field using the \synthmag\ code \citep{piskunov99}. The line of Cr\,\textsc{ii} 5116\,\AA\ with a large Land\'e factor of $ g = 2.92 $ is rather weak and asymmetric with a stronger blue part of the profile in the first observing set of 2004 and with a stronger red part for the second observing set from 2006. The synthetic line profile fit to this line using \synthmag\ needed no magnetic field strength greater than $1$--$1.5$\,kG. Line profiles with large Land\'e factors belonging to iron, such as Fe\,\textsc{i} at 5501\,\AA, 5507\,\AA\, and 6337\,\AA, fitted slightly better for a magnetic field strength up to $2$\,kG, in comparison with synthetic spectra with zero magnetic field. We therefore place an upper limit of $\langle H \rangle < 2$\,kG. 

\cite{romanyuk04} compared longitudinal magnetic field extrema with the mean magnetic field modulus for 39 Ap stars and found a linear relation:  $\langle H \rangle = 1.01 + 3.16 \langle H_l \rangle$\,kG. Using this relation and the extremum of the longitudinal field measurements obtained by \cite{borra75} and \cite{mathyshubrig97} gives a mean magnetic field modulus of 1.8\,kG and 1.3\,kG, respectively. These values are consistent with the upper limit estimated from our UVES spectra. 
 
\subsection{Simplifications: magnetic fields and NLTE\label{sec:simp}}

All atmospheric models include simplifications 
to the physical conditions in real atmospheres. 
This is especially the case for the roAp stars, in which magnetic fields 
may alter the atmospheric structure \citep{koch05} and where observational evidence
for stratification of certain elements is found \citep{ryab02}.
More realistic models of magnetic stars are being developed \citep{koch05},
but they indicate that the structure in \acir\ should only be
mildly affected by the relatively weak magnetic field of
$\le 2$\,kG that we determined above.

The \atlasni\ models we applied assume local thermodynamical equilibrium (LTE), 
but at \teff\ above $7000$~K non-LTE effects become important.
For a normal star with solar metallicity and \teff\,$=7500$\,K, the correction for neutral iron is
[Fe\ione/H]$_{\rm NLTE}=$\,[Fe\ione/H]$_{\rm LTE}+0.1$\,dex \citep{rent96}.
This seems to agree with the slight difference in Fe abundance we find from neutral and ionized lines.
We adopt this correction for our final result for the photospheric metallicity, which is
the mean of the 125 Fe\ione\ lines in \acir\ and, hence, \feh\,$=-0.12\pm0.08$.
We note that the overall metallicity of \acir, below its photosphere, is much more uncertain.


%
%
\section{Conclusion}

We present the first angular diameter measurement of an roAp star, \acir,
based on interferometric measurements with SUSI.
In combination with our estimate for the bolometric flux and 
the new \hipp\ parallax, we have determined the stellar radius 
and effective temperature nearly independent of theoretical atmospheric models.
The constraints found in this work on the luminosity, \teff\ and radius
will be invaluable for critically testing theoretical pulsation models of \acir.
We will present this in our companion paper, 
where we report the detection of the large frequency spacing in \acir, 
based on photometry from the \wire\ satellite \citep{bruntt08a}.

We analysed high-resolution spectra taken at two different rotational phases.
The detailed abundance analyses yield the same mean abundances for the two spectra,
but differences in the line shapes of certain elements 
give evidence for spots on the surface of \acir.
We confirm the abundance pattern seen in \acir\ in an earlier
study using fewer lines \citep{kupka96}. 
The results follow the general trend in other roAp stars, 
\ie\ high photospheric abundance of Cr, Co, Y and the rare-earth elements Nd and Eu.


%
%

\section*{Acknowledgments}

This research has been jointly funded by the University of Sydney and the 
Australian Research Council as part of the Sydney University Stellar
Interferometer (SUSI) project. 
We acknowledge the support provided by a University of Sydney
Postgraduate Award (APJ) and Denison Postgraduate Awards (APJ and SMO).
VGE and DWK acknowledge support from the UK STFC.
MC is supported by the European Community's FP6, FCT and FEDER (POCI2010) 
through the HELAS international collaboration and through the projects 
POCI/CTE-AST/57610/2004 and PTDC/CTE-AST/66181/2006 FCT-Portugal.
This research has made use of the SIMBAD database,
operated at CDS, Strasbourg, France.
We made use of atomic data compiled in the VALD database \citep{vald}.
We thank Christian St\"utz for supplying the \llmodels\ \citep{shulyak04} 
we used in the analysis.
We wish to thank Brendon Brewer for his assistance with the theory and 
practicalities of Markov chain Monte Carlo simulations.

%
%
 
\bibliography{bruntt_acir_parameters_astroph}

\bsp
\label{lastpage}

\end{document}